# John Goodricke, Edward Pigott, and Their Study of Variable Stars

**Linda M. French**

*Illinois Wesleyan University, Department of Physics, P.O. Box 2900, Bloomington, IL 61702; lfrench@iwu.edu*

**Abstract**   John Goodricke and Edward Pigott, working in York, England, between 1781 and 1786, determined the periods of variation of eclipsing binaries such as β Persei (Algol) and β Lyrae and speculated that the eclipses of Algol might be caused by a "dark body," perhaps even a planet.  They also determined the periods of variation of the first two known Cepheid variables, the stars whose period-luminosity relation today enables astronomers to determine distances to distant galaxies.  Goodricke holds special interest because he was completely deaf and because he died at the age of 21.  The lives and work of these two astronomers are described.

## 1. Introduction

The name of John Goodricke (1764–1786; Figure 1) is recognized by many astronomers today, but few details of his life and work are widely known. Some know that he observed variable stars, some know that he was profoundly deaf, and some know him as an amateur astronomer. Goodricke's collaborator, Edward Pigott (1753–1825), is even less well known. Together, these two determined the periods of variation of eclipsing binaries such as β Per (Algol) and β Lyrae, speculating that the eclipses of Algol might be caused by a "dark body," perhaps even a planet. They also discovered and determined the periods of variation of η Aquilae and δ Cephei, the first two known Cepheid variables. The period-luminosity relation of Cepheids, of course, would later enable astronomers to determine distances to distant galaxies. In 2010, the author was able to spend a sabbatical semester at the University of York, studying the journals and notebooks of Goodricke and Pigott in order to understand how these pioneers went about their work.

Richard Holmes (2008) cautions about the shroud of myths that often envelops scientists of great accomplishment. One such myth is that of the lone, heroic figure, struggling against misconceptions perpetrated by lesser minds, against his (or her) own family, and perhaps even against society itself. This myth does not apply to John Goodricke. Rather, he was able to attend forward-thinking schools that addressed his learning needs and nurtured his talents, and he had the support of a family who clearly valued and encouraged his studies.

## 2. John Goodricke: background and family life

The Goodricke family line is long, with several branches in England. The Goodrickes of Yorkshire took up residence at Ribston, just west of the city of York, in 1533 when Henry Goodricke became steward of Great Ribston (Figures 2 and 3). In 1641 Sir John Goodricke was created the first Goodricke baronet for his service to the King during the Civil Wars. John, the astronomer, was the eldest grandson of the fifth baronet, also named Sir John (1708–1789).

Zdeněk Kopal, the Czech-British astrophysicist, once described the Goodrickes as "fox-hunting country squires" (Kopal 1986), a characterization that the facts do not support. The Goodricke baronets were, for the most part, not content to sit at home on the Ribston estate. Sir Henry, the second baronet, was the English Ambassador to Spain from 1678 until 1682, and Sir John, the astronomer's grandfather, was Envoy Extraordinary to Sweden from 1764 until 1773. Both men, as well as the astronomer's father, Henry (1741–1784), served as Members of Parliament, and both baronets were members of the King's Privy Council (somewhat similar to the U.S. President's Cabinet).

John Goodricke, the astronomer, was born on 17 September 1764 in Groningen, the Netherlands, where his father, Henry, was employed in diplomatic service. John's mother was born Levina Benjamina Sessler; her father was Peter Sessler, a merchant of Namur, Belgium. John was the eldest surviving child, and so he would have been the heir to the baronetcy had he lived to succeed his father and grandfather.

According to the family history, John became deaf at the age of five due to a severe illness that has been conjectured to be scarlet fever. At the age of seven he went to study at Thomas Braidwood's Academy for the Deaf and Dumb in Edinburgh, the first school for deaf children in the British Isles. Braidwood was very secretive in his teaching methods. We do know that Braidwood advertised "to undertake to teach anyone of a tolerable genius in the space of about three years to speak and to read distinctly" (quoted in Pritchard 1963); that his pupils read lips and signed; and that Braidwood had originally been a mathematics teacher (Branson and Miller 2002).

John went on to study at the Warrington Academy for three years after leaving Braidwood's. Warrington was one of the "Freethinking" or Non-Conformist academies originally founded to prepare clergymen in denominations other than the Church of England. It was well known for its emphasis upon mathematics and natural philosophy; the chemist Joseph Priestley had taught there but had moved on before John Goodricke arrived in 1778 (Parker 1914). John was described as "a very tolerable classic and an excellent mathematician" in a school report (Turner 1813). During John's time at Warrington, the mathematics curriculum (which included astronomy in the second year) was taught by William Enfield (McLachlan 1943). Enfield was primarily a theologian, but he worked diligently at his teaching and eventually published his notes as a textbook, *Institutes of Natural Philosophy* (Enfield 1785), which went through many editions on both sides of the Atlantic. John's mathematics notebook is preserved in the Goodricke collection of the York City Archives, and the figure seen there can be found on the inside back cover (Goodricke 1779; Figure 4).

## 2.1. The Warrington sketch

The drawing shows several constellations: Orion's belt can be seen, along with the brightest star in Taurus, "The Eye of the Bull," Aldebaran; the constellation of Auriga; and "the two brightest stars in the Gemini." The Milky Way is shown, as well as the zodiac (or ecliptic), and the Moon. At the bottom of the page is a sentence describing the position of various stars on either side of the meridian, a line connecting the north and south points on the horizon and passing through the zenith. The star positions, together with the Moon's position in the sky, permit determination of the approximate date of the drawing. The only time that matches both the Moon and star positions is a one-week period in late November of 1779. On 23 November 1779, a total lunar eclipse was visible over England (Borkowski 1990). John Goodricke would have had access to textbooks with tables of predicted eclipses (such as Ferguson 1756); he would also have been taught to do such calculations in his schoolwork (Enfield 1785). Exactly how he came to produce this drawing we may never know. What is significant, however, is that he was already observing the sky in 1779, at the age of fifteen.

## 2.2. Correcting some popular misconceptions about John Goodricke: a "deaf-mute"?

John Goodricke is often described as "deaf and dumb," or a "deaf-mute." Evidence from the Goodricke journals suggests that, while he was certainly deaf, he almost certainly spoke. He evidently was able to read lips (teaching students to lip-read and to speak if they were capable of it was part of the curriculum at the Braidwood Academy). The evidence for this is in two passages from Goodricke's *Journal of the Going of My Clock* (Goodricke 1782a):

> 17 November 1782: Whilst I was winding up the Clock the second hand did not go on as usual–I spoke to Mr Hartley [the clockmaker] about it & he said it was caused by my not pulling down the Spring hard enough....

> 15 December 1782: Whilst I was winding up the Clock on the 15th the second hand did not go on as usual–As this is now the 3rd time it did so; I remonstrated with Mr Hartley about it & asked him ye reasons of it doing so–He gave me the same answer as on the 17th of Nov. last but I did not credit him–However after several trials I have since hit upon the true course & found that it was owing to a fault of my own in not pulling the spring down hard enough according to
Hartley's directions which I did not rightly understand or he was not
very particular in explaining them to me.

From the words alone, nothing could be clearer: he *spoke* with Mr. Hartley,
he *remonstrated* with Mr. Hartley. The second passage makes it even more
explicit that the conversation was a verbal one; Hartley explained and Goodricke
did not initially understand the explanations. Had the directions been written
out, it is much less likely that such a breakdown in communication would have
occurred. Thus, the available evidence suggests that Goodricke read lips well
enough to carry on business transactions, and that he may well have spoken.

### 2.3. Burial Place

Zdeněk Kopal, in his scientific autobiography *Of Stars and Men* (1986),
described a visit to the churchyard of St. John the Baptist at Hunsingore
(Figure 5), the burial place of John and the other Goodrickes, and came to the
conclusion that John Goodricke had been buried apart from his family in an
unmarked grave. Kopal wrote: "Why does he rest there forgotten by all his
clan; why was he not buried with them in their family vault[?]...." He went on
to speculate that John's parents and grandparents found his deafness to be "a
blot on the family escutcheon." Kopal apparently did not investigate the history
of the present church; if he had he would have discovered that it dates to 1868,
after the Goodricke family estate at Ribston had been purchased by the Dent
family. There was a Goodricke family vault under the old church, and that vault
still exists. It is marked in the churchyard by a stone identical to that used for
the new church, with only the words "Goodricke Vault" engraved upon the side
(Figure 6). The burial records still exist (N. Yorkshire County Record Office
MIC 1685), and they show that John Goodricke was indeed buried alongside
his parents and grandparents in the family vault. Although the deaf were often
treated inhumanely in the eighteenth century, John Goodricke's family gave him
the best possible education both for his scientific research and for his stature as
the Heir Apparent to a baronetcy.

The previous Goodricke baronets had attended university at either
Cambridge or Aberdeen, and John surely would have been intellectually
qualified for university. Why he returned to York at seventeen to live with his
family is somewhat puzzling. Both John, in his journal, and Edward Pigott, in a
diary, make occasional references to John's not being well, so perhaps his health
had already begun to fail. At any rate, the first entry in John's formal observing
journal (Goodricke 1781) comes early November 1781, when he writes: "Last
evening at 9 p.m. Mr. E. Pigott discovered a comet."

During the first few entries John describes Edward's correspondence with
William Herschel and with Nevil Maskelyne, the Astronomer Royal. Edward's
contacts in the astronomical world, as well as his discoveries, clearly impressed
John, who immediately set about keeping a record of his own observations.

## 3. Edward Pigott: background and family life

Edward Pigott's father, Nathaniel (1725–1804), was also an astronomer, and he was the primary source of Edward's astronomical training. The Pigotts were related to the wealthy, landed Fairfax family of Yorkshire; Nathaniel's mother Althea Fairfax Pigott was the sister of Charles Gregory Pigott (d. 1772), ninth Lord Fairfax and Viscount Emley. As Catholics, the Pigotts found life in France more congenial than life in the north of England, and they spent a great deal of time there. Edward went to school in both countries, but French was his first language, which gives an occasional "invented" feel to the wording and spelling of his journals.

Nathaniel's primary interest was in using astronomical methods such as the timing of eclipses of the Moon and the Jovian satellites to determine latitude and longitude. Although not a wealthy man, he was able to acquire instruments made by the finest craftsmen of the time, including Ramsden, Dollond, Sisson, and Bird. Between 1773 and 1775 Nathaniel and Edward collaborated with continental astronomers including Messier and Mechain to determine the latitude and longitude of several cities in the Austrian Netherlands (now Belgium; Pigott 1778).

Nathaniel Pigott owned property in Middlesex and in Wales, and in 1781 the family settled in York, where Nathaniel hoped to manage the estates of Lady Anne Fairfax, the sole surviving daughter of Lord Fairfax, and to eventually secure the estates as an inheritance for Edward's younger brother, Charles Gregory Pigott. The Pigott family took up residence in York, approximately one-quarter mile from where the Goodrickes were living. Here Nathaniel constructed an observatory said to be amongst the finest private observatories in England.

A diary kept primarily by Edward Pigott with some entries by Nathaniel (now in the Beinecke Library of Yale University) includes stories of joint Goodricke-Pigott family outings. Thus, even though the start of the official collaboration dates from John's beginning to keep the observing journal, it seems likely that the two discussed astronomy at an earlier date.

### 3.1. Interest in variable stars

Stellar astronomy was still in its infancy in the eighteenth century (see, for example: Hoskin 1982; Williams and Hoskin 1983). Among variable stars, a period had been determined only for the long period variable o Ceti (Mira). Ismael Boulliau, better known by his Latinized name Bullialdus, observed the star systematically between 1660 and 1666, obtaining an accurate period of nearly 333 days (Hoskin 1982; Hatch 2011). Bouilliau went on to consider sources of the star's variability, and hypothesized that the most likely cause of the variation was dark regions on the star coming into view as it rotated; in other words, spots analogous to sunspots. That long period variables do not always show an exact periodicity or reach the same peak brightness was to be expected, since the variation in the Sun's light due to sunspots is not exact. Boulliau's explanation was accepted and adopted by Newton in Book 3 of the *Principia*, and by William Herschel in his first published paper (1780), which contained observations of Mira (Hatch 2011).

As early as 1778, while observing from Wales, Edward Pigott was noticing that both the reported positions and brightnesses of stars varied from one catalogue to another, and he speculated on possible sources of the noted discrepancies. He continued this practice from York. In July of 1781, for example, Edward wrote in his journal:

> The 22nd star of Tycho's Andromeda is probably the o (omicron) of that constellation, tho' it differs very considerably both in Longitude and Latitude, which I am convinced is occasioned by an error either in the Observation or Calculation, the Prince Hesse [probably William IV, Landgrave of Hesse-Kassel] observed the o therefore it was visible in Tycho's times and has been since; See Hevelius's & Flamsteed's Observations; now it is not probable that Tycho would have overlooked a star of the 3rd or 4th mag. (Pigott 1781)

A discussion of the positional uncertainties in the catalogs of Tycho, Hevelius, Flamsteed, and the Landgrave is beyond the scope of this paper. What is significant in this passage is Edward's taking note of discrepant magnitude estimates and commenting that Tycho would not have omitted a star as bright as the third or fourth magnitude—exactly the magnitude range of the stars that he and John would soon study systematically. The implication is that the star might well have varied in brightness.

In the autumn of 1782 John and Edward decided to pursue observations of "Stars which are Variable or Thought to be so," as John wrote in the heading of one journal entry in early November (Goodricke 1782b). The first star on his list is β Persei (Algol), whose changes in brightness had been noted as early as 1672 by the Italian astronomer Geminiano Montanari. In October 1782, Edward Pigott noted, "This star is variable" for Algol, almost certainly as a result of a literature search, as he had made no extensive observations of the star up to that date. Other stars on John's list as candidates for variability included δ Ursae Majoris, not thought today to be variable, and α Herculis, now classed as a semiregular variable with amplitude of nearly one magnitude.

On 12 November 1782, John noted,

> This night I looked at Beta Persei [Algol], and was much amazed to find its brightness altered—It now appears to be of about 4th magnitude. I observed it diligently for about an hour—I hardly believed that it changed its brightness because I never heard of any star varying so quickly in its brightness. I thought it might perhaps
be owing to an optical illusion, a defect in my eyes, or bad air, but the
sequel will show that its change is true and that I was not mistaken.
(Goodricke 1782c)

   The two began checking Algol every clear night. They did not see another
diminution of light until 28 December. By April they had seen consecutive
episodes of darkening, and were able to determine that the period was very
short compared to that of Mira: only 2 days and 21 hours. According to the
custom of the time for reporting scientific results, John sent off a memorandum
to Anthony Shepherd, Plumian Professor of Astronomy at Cambridge, to be
read at the Royal Society of London. At the same time, Edward Pigott notified
Nevil Maskelyne, the Astronomer Royal, and William Herschel, both of whom
were eager to observe Algol. The variability was quickly confirmed by Herschel
and other astronomers of the Royal Society. In his report, published in the
*Philosophical Transactions of the Royal Society*, John states:

   I should imagine [the diminution of light] could hardly be accounted
   for otherwise than either by the interposition of a large body revolving
   round Algol, or some kind of motion of its own, whereby part of its
   body, covered with spots or such like matter, is periodically turned
   towards the earth. (Goodricke 1783)

   The two discussed the idea of a "large body" revolving around Algol,
as their journals both indicate, and in the journals both call the large body
a planet. It is likely, as Michael Hoskin (1982) suggests, that the planet
hypothesis originated with Edward Pigott, the more experienced observer and
always the more adventurous theorizer of the two. Yet Goodricke wrote the
formal report, and in August of 1783 he was awarded the Copley Medal of the
Royal Society.
   We now believe transits of a fainter stellar companion to be the correct
explanation for the Algol system. Observations of transits are currently being
used by NASA's Kepler mission to detect Earthlike planets around other stars.
Yet in their own time Goodricke and, to a lesser extent Pigott, would abandon
the transit hypothesis in favor of starspots. In his last completed paper, on the
period of variation of δ Cephei, Goodricke would write:

   What I have before mentioned, that the greatest brightness of δ
   Cephei does not seem to be always quite the same, is not peculiar
   to this star, but is also to be observed in the other variable ones....
   Even Algol does not seem to be always obscured in the same degree,
   being perceived to be sometimes a little brighter than ρ Persei, and
   sometimes less than it....This may, I suppose, be accounted for by a rotation of the star on its axis, having fixed spots that vary only in their size. (Goodricke 1786)

Several factors could have contributed to Goodricke's change of mind. By this time, he had visited Nevil Maskelyne at Greenwich and been exposed to the opinions of senior astronomers, who favored sunspots, as we have seen. But also, the nature of δ Cephei's light curve differs from that of Algol. There is not one single isolated diminution, but a continuous fading and brightening; a pattern that is less easily interpreted in terms of an eclipse. Finally, ρ Persei, conveniently placed for comparison with Algol, is itself a variable star, and so it may well have been "sometimes a little brighter" and sometimes less bright than Algol. Most modern observers can relate to the dilemma of choosing a comparison star that turns out to be variable! Only a century later was the eclipse hypothesis confirmed using spectral analysis (see Batten 1989 for a review).

## 4. Other astronomical work

John Goodricke's remaining time on Earth was short. He continued to observe Algol; in addition to determining the period of δ Cephei he also obtained the period of β Lyrae. In the autumn of 1784, as Goodricke studied δ Cephei, Edward Pigott detected the variation of another Cepheid, η Antinoi (today η Aquilae). Edward would eventually discover two more variable stars, R Scuti and R Coronae Borealis; he discovered the spiral galaxy known as M64 before Bode, and Jerome La Lande would write him that

The observations which you sent me in 1782…have been very useful in my research into a theory for Mercury, which I have published… their ephemerides showed me for the first time that the place of the aphelion was too far advanced in my tables. (LaLande 1786)

Thus, Edward Pigott's observations may well have been among the first showing the advance of the perihelion of Mercury!

John Goodricke died on April 20, 1786, in York, 14 days after being elected to membership in the Royal Society at the age of 21. Edward Pigott completed their determination of the latitude and longitude of York and wrote of Goodricke:

This worthy young man exists no more; he is not only regretted by many friends, but will prove a loss to astronomy, as the discoveries he so rapidly made sufficiently evince: also his quickness in the study of mathematics was well known to several persons eminent in that line. (Pigott 1786)

## 5. The Goodricke-Pigott legacy

John Goodricke is better known today than Edward Pigott. The University of York has a Goodricke College, and the dramatic story of Goodricke's short life figures prominently in several astronomical textbooks (for example, Fraknoi *et al.* 2006). Surely Goodricke's being awarded the Copley Medal and elected to membership in the Royal Society brought him recognition. It is clear that Edward Pigott deserves at least equal credit for their joint work. Today, Edward would be recognized as a co-discoverer of the periods of Algol, δ Cephei, and β Lyrae, while John would be credited with helping discover the period of η Aquilae and determining the coordinates of York.

The petition nominating John Goodricke to membership in the Royal Society was apparently initiated by Nathaniel Pigott; co-signers include Nevil Maskelyne, Anthony Shepard, Thomas Hornsby, Savilian Professor of Astronomy at Oxford, and William Wales, a member of the Board of Longitude, among others. Edward Pigott, on the other hand, although deserving, was never even nominated. Was this due to differences in the social standing of the two? Was there a reluctance on Nathaniel's part to push for his son's nomination? Or was Edward simply not considered a "clubbable man"? It is possible that all of these played a part.

What is certain is that the two held each other in high regard and frequently expressed that regard both in their journals and in their publications. Edward Pigott felt, justly, that his father Nathaniel did not give him enough credit for his astronomical work, and it is certain that Nathaniel cut Edward out of his will, as evidenced by Edward's pleading letters to his great-aunt Lady Anne Fairfax (N. Yorkshire County Record Office ZDV F: MIC 1132/1201). Edward did not suffer slights lightly. Yet Edward frequently mentions John Goodricke's talents both as an observer and in the interpretation of data. Neither in print nor in Edward's journals is there any hint that he resented Goodricke's authorship of the Algol paper, his reception of the Copley medal, or his election to the Royal Society.

John Goodricke clearly admired and learned from Edward Pigott. Edward's long-held interest in the nature of the stars, especially their possible variability, flowered into a productive scientific research program almost as soon as he and John Goodricke began their joint investigations. These two deserve to be better known, and to share joint credit for their discoveries.

## 6. Acknowledgements

It is a pleasure to thank Joy Cann and Caroline Stockdale for assistance while working at the York City Archives. The author thanks Alison Brech, Charles and Annie Dent, Martin Lunn, Anita McConnell, John Percy, and Ian Stuart for fruitful discussions; Ron Emmons for research assistance; and Thomas

Williams for a helpful referee's report. Research for this work was carried out while the author was a Visiting Professor of Physics in the University of York. Support from the Herbert C. Pollock Award of the Dudley Observatory and from a Small Research Grant of the American Astronomical Society is gratefully acknowledged.

Unless otherwise cited, information on John Goodricke and his family comes from a family history originally written by Charles Alfred Goodricke (1897). An abbreviated version is currently maintained online by Michael Goodrick (2010). The primary source of information on Edward Pigott is the 1999 article by Anita McConnell and Alison Brech (1999) entitled "Nathaniel and Edward Pigott, Itinerant Astronomers."

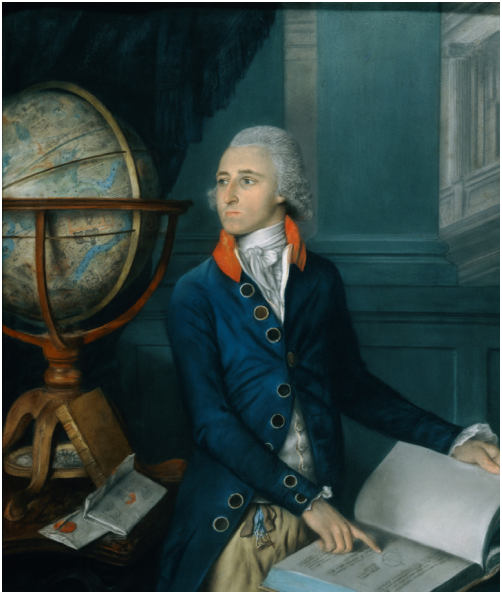

Figure 1. John Goodricke (1764–1786). Pastel portrait by James Scouler, now the property of the Royal Astronomical Society. Used with permission of the RAS.

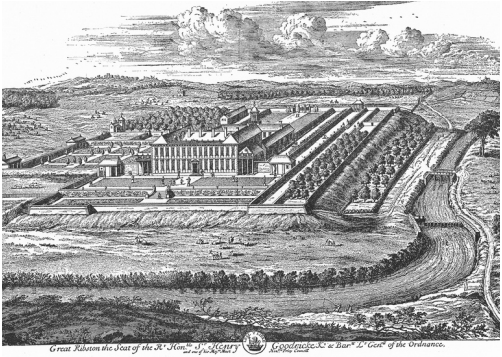

Figure 2. Ribston Hall in the seventeenth century. From the Goodricke family history website maintained by Michael Goodricke at http://www.goodrick.info/main.htm

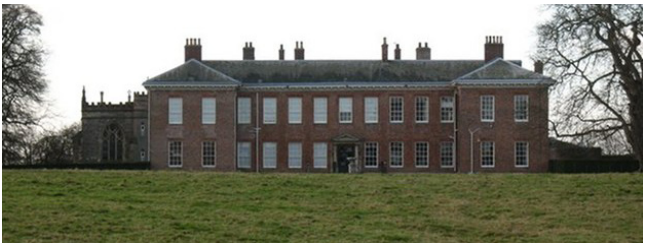

Figure 3. Ribston Hall today. © Copyright Gordon Hatton <http://www.geograph.org.uk/profile/4820> and licensed for reuse under this Creative Commons License<http://creativecommons.org/licenses/by-sa/2.0/

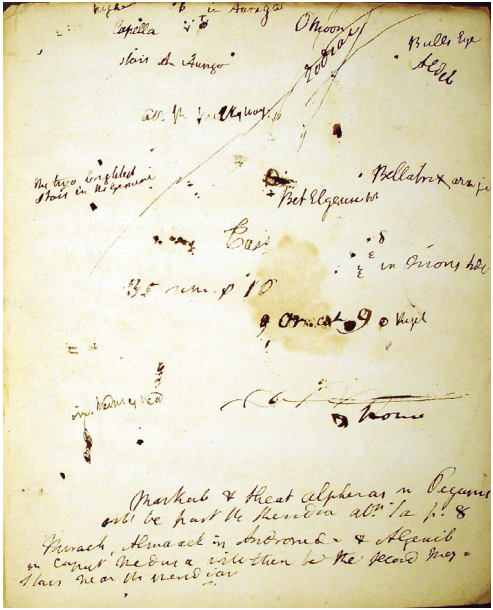

Figure 4. Drawing found in the inside back cover of John Goodricke's mathematics notebook from Warrington Academy, 1779–1780. The constellations of Orion, Taurus, Auriga, and Gemini are shown, along with the Moon, Milky Way, and Zodiac. Positions of stars are given that are consistent with the drawing having been made in November 1779. Reproduced from an original held by City of York Council Archives and Local History (Goodricke 1779).

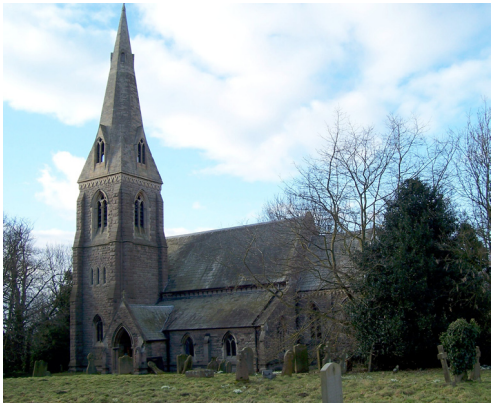

Figure 5. The church of St. John the Baptist in Hunsingore. The low, flat stone just to the left of center in the photograph marks the location of the Goodricke vault.

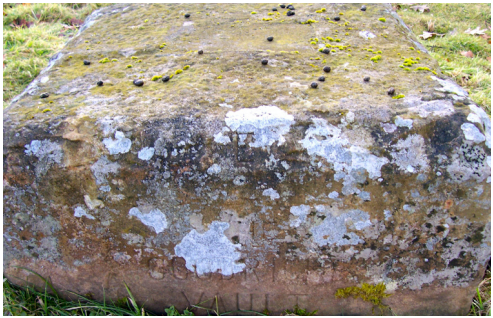

Figure 6. The east-facing side of the marker stone for the vault. The only engravings are the letter "E" at the top and the words, "The Goodricke Vault."